\documentclass{article}

\usepackage{PRIMEarxiv}

\usepackage[utf8]{inputenc} 
\usepackage[T1]{fontenc}    
\usepackage{hyperref}       
\usepackage{url}            
\usepackage{booktabs}       
\usepackage{amsfonts}       
\usepackage{nicefrac}       
\usepackage{microtype}      
\usepackage{lipsum}
\usepackage{fancyhdr}       
\usepackage{graphicx}       
\usepackage{subcaption}
\usepackage{caption}

\graphicspath{{media/}}     

\title{Enhancing Diagnosis through AI-driven Analysis of Reflectance Confocal Microscopy
\thanks{This manuscript has been authored in part by UT-Battelle, LLC, under contract DE-AC05-00OR22725 with the US Department of Energy (DOE). The US government retains and the publisher, by accepting the article for publication, acknowledges that the US government retains a nonexclusive, paid-up, irrevocable, worldwide license to publish or reproduce the published form of this manuscript, or allow others to do so, for US government purposes. DOE will provide public access to these results of federally sponsored research in accordance with the DOE Public Access Plan (http://energy.gov/downloads/doe-public-access-plan).}
} 

\author{
  Hong-Jun Yoon\\
  Oak Ridge National Laboratory \\
  Oak Ridge, TN USA\\
  \texttt{yoonh@ornl.gov} \\
   \And
  Chris Keum\\
  Farragut High School \\
  Knoxville, TN USA\\
   \And
  Alexander Witkowski\\
  Oregon Health and Science University \\
  Portland, OR USA\\
   \And
  Joanna Ludzik\\
  Oregon Health and Science University \\
  Portland, OR USA\\
   \And
  Tracy Petrie\\
  Oregon Health and Science University \\
  Portland, OR USA\\
   \And
  Heidi A. Hanson\\
  Oak Ridge National Laboratory \\
  Oak Ridge, TN USA\\
   \And
  Sancy A. Leachman\\
  Oregon Health and Science University \\
  Portland, OR USA\\
}

\begin{document}
\maketitle

\begin{abstract}
Reflectance Confocal Microscopy (RCM) is a non-invasive imaging technique used in biomedical research and clinical dermatology. It provides virtual high-resolution images of the skin and superficial tissues, reducing the need for physical biopsies. RCM employs a laser light source to illuminate the tissue, capturing the reflected light to generate detailed images of microscopic structures at various depths. Recent studies explored AI and machine learning, particularly CNNs, for analyzing RCM images. Our study proposes a segmentation strategy based on textural features to identify clinically significant regions, empowering dermatologists in effective image interpretation and boosting diagnostic confidence. This approach promises to advance dermatological diagnosis and treatment.

\end{abstract}

\keywords{reflectance confocal microscopy \and skin cancer \and dermatology \and vision transformer \and self supervised learning \and artificial intelligence}

\section{Introduction}
Reflectance Confocal Microscopy (RCM) marks a paradigm shift in biomedical imaging, offering a sophisticated, non-invasive technique to acquire high-resolution images of the skin and superficial tissues. Its development \cite{calzavara2008reflectance} represents a milestone in medical imaging, transitioning from early exploratory stages to becoming a cornerstone in clinical dermatology. RCM's capability for in vivo imaging, capturing live tissue images without the need for biopsies or tissue excision, has made it an indispensable tool in modern medical diagnostics. The inception of RCM can be traced back to its early conceptualization, where the need for less invasive, more accurate diagnostic methods in dermatology was recognized. Over the years, the technology has undergone significant advancements, evolving in its design and functionality. This evolution has been marked by improvements in laser source quality, detector sensitivity, and image processing algorithms, resulting in enhanced image clarity and depth of tissue analysis. RCM's operation relies on a focused laser light to illuminate the target tissue. The tissue interaction with this light, primarily through backscattering and reflection, forms the basis of image creation. A detector captures this reflected light, and sophisticated computer algorithms transform these signals into detailed, cross-sectional images of the tissue. This allows for the visualization of cellular and sub-cellular structures with remarkable clarity and precision.

In dermatology, RCM has proven particularly beneficial. It provides a real-time, non-invasive alternative to traditional skin biopsies, allowing for immediate assessment of skin lesions. Conditions like melanoma, basal cell carcinoma, and squamous cell carcinoma can be diagnosed with greater accuracy, significantly improving patient outcomes \cite{levine2018introduction}. Moreover, RCM facilitates the monitoring of treatment responses and the management of chronic conditions, offering a dynamic tool for personalized patient care. Comparatively, RCM holds distinct advantages over other imaging modalities such as ultrasound, magnetic resonance imaging (MRI), or traditional dermoscopy. Its higher resolution and ability to image at a cellular level give it an edge in early disease detection and diagnosis, particularly in cases where other modalities may fall short.

In recent years, the integration of AI and ML technologies, especially convolutional neural networks (CNNs), has further expanded the capabilities of RCM \cite{wodzinski2019convolutional,sikorska2021learning,campanella2022deep,patel2023analysis}. These technologies enable the processing and interpretation of complex image data, facilitating automated detection and diagnosis of skin pathologies. This synergy of RCM and AI presents an exciting frontier in dermatological imaging, offering prospects for more accurate, efficient, and personalized patient care. Our study contributes to this evolving landscape by proposing a novel segmentation strategy for RCM images, focusing on textural features to identify key clinical regions. This approach aims to augment the diagnostic accuracy and efficiency, potentially transforming how dermatologists interpret RCM images and make clinical decisions. 


Looking forward, the potential of RCM extends beyond dermatology. Its application could revolutionize areas such as ophthalmology, neurology, and oncology, where non-invasive imaging plays a critical role. Furthermore, the ongoing integration of RCM with emerging technologies such as augmented reality, 3D imaging, and telemedicine could further enhance its utility, making it a versatile tool in various medical disciplines. The future of RCM is poised to be marked by continuous innovation, with the promise of significantly impacting patient care and medical research.

\section{Methods}
\subsection{Dataset}

The dataset utilized in this study comprised a total of 519 de-identified Reflectance Confocal Microscopy (RCM) images provided by the esteemed Department of Dermatology at Oregon Health and Science University. The dataset was balanced, consisting of 233 images that were recommended for biopsy and 286 images that were not recommended for biopsy. 
This collection is composed of various skin conditions of varying cancer likelihoods.
Note that the RCM images included in the dataset exhibit two pixel resolutions. Specifically, a subset of the images possesses a resolution of 1,000 by 1,000 pixels, whereas the remaining images have a resolution of 1,024 by 1,024 pixels. All the images are rendered in single-channel 8-bit grayscale. All images were de-identified to ensure patient confidentiality, with any personal identifiers removed.

\subsection{Image patch generation}

We encountered a challenge with the image sizes of the RCM images in the dataset, as they exceeded the commonly used de-facto standard resolution of 256 by 256 pixels in AI/ML models for image processing and classification. In order to align with the current standards in AI/ML, we made the decision to divide each RCM image into multiple patches, each measuring 256x256 pixels. This division allowed us to maintain compatibility with AI/ML models while exploring the intricate characteristics of the images. Furthermore, this approach aligns with the primary objective of our study, which is to investigate image characteristics using computer algorithms and enable automatic segmentation with minimal expert intervention.

\begin{figure}
    \centering
    \begin{subfigure}[t]{0.47\textwidth}
        \centering
        \includegraphics[width=0.95\linewidth]{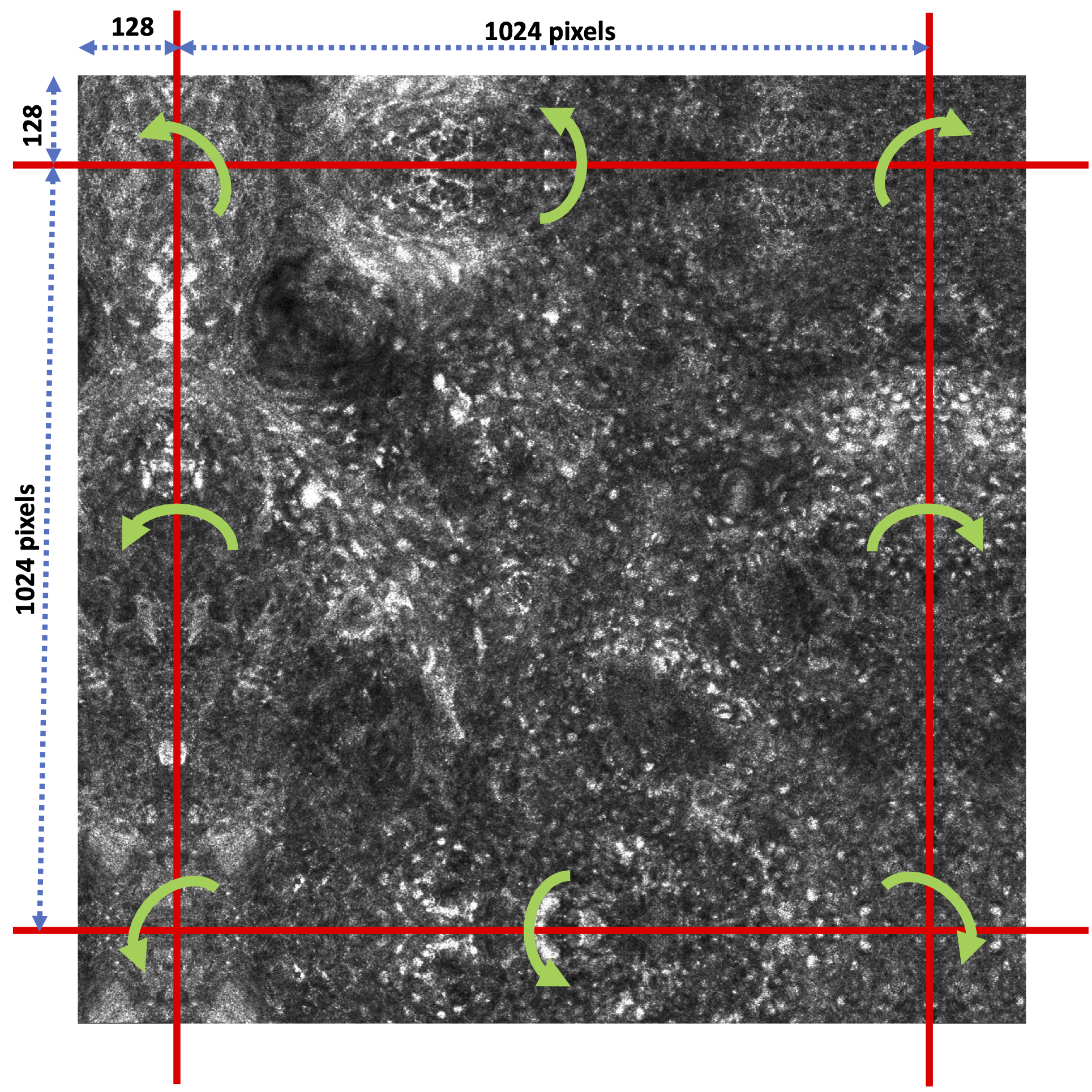}
        \caption{}
    \end{subfigure}
    \begin{subfigure}[t]{0.47\textwidth}
        \centering
        \includegraphics[width=0.85\linewidth]{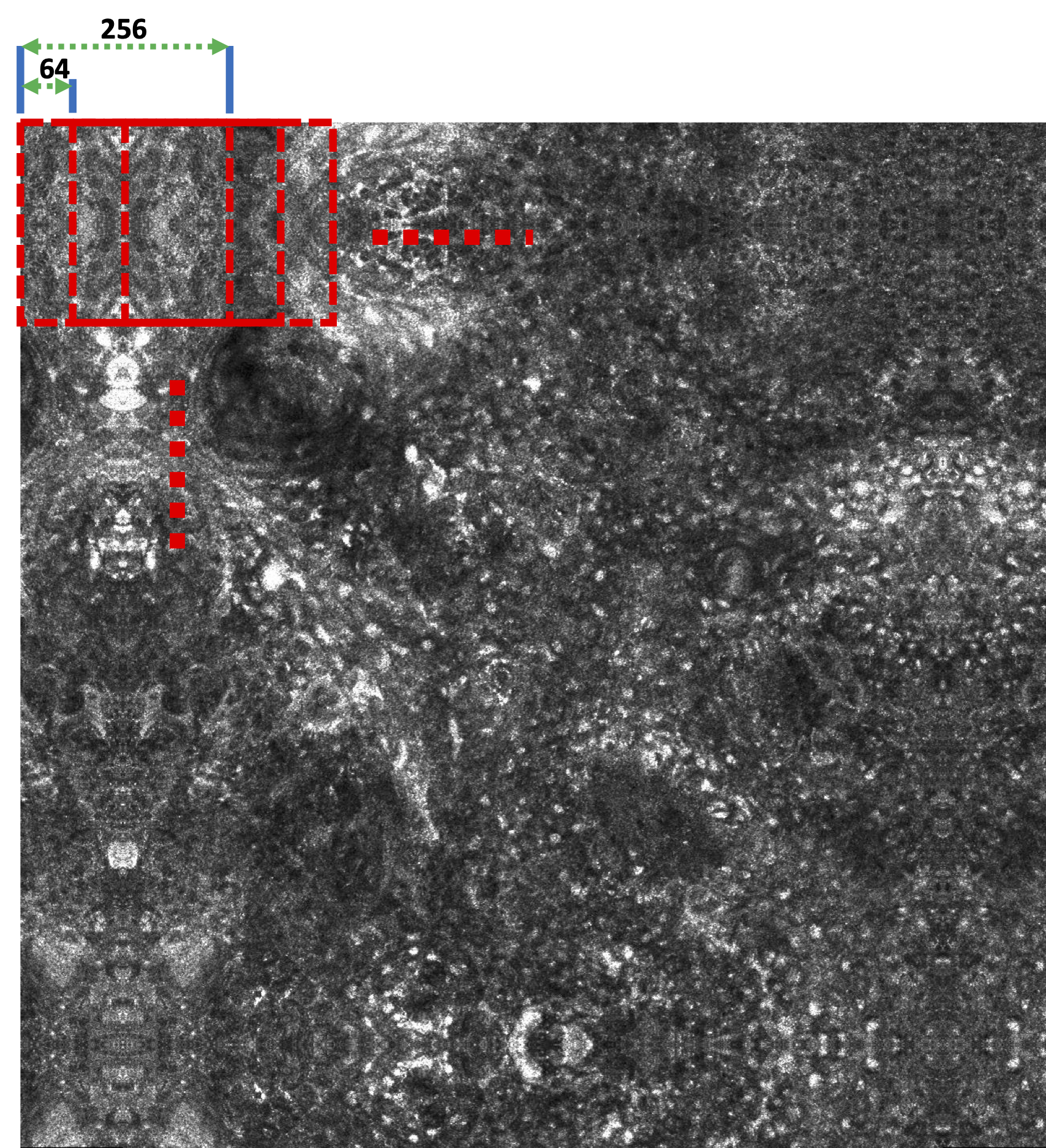}
        \caption{}
    \end{subfigure}
    \caption{Patch generation from RCM images. (a) applying mirroring to the edges of image and (b) making image patches of 256x256 pixels.}
    \label{fig_patch}
\end{figure}

Recognizing the significance of boundary information within these patches, we implemented a step size of 64 pixels between adjacent patches. This step size was chosen to preserve the boundary characteristics and fine details within the images. In order to ensure comprehensive coverage of the image space, we employed a mirroring technique to extend the boundaries of the RCM images by 128 pixels. This allowed for a thorough exploration of image boundaries. Consequently, a single RCM image with a resolution of 1,000 by 1,000 pixels resulted in the generation of 256 distinct image patches. Similarly, an image with a slightly higher resolution of 1,024 by 1,024 pixels produced 289 patches. Figure \ref{fig_patch} illustrates the procedure of mirroring and image patch creation from the RCM images.

\subsection{Feature extraction}

\begin{figure}
    \centering
    \includegraphics[width=1.0\linewidth]{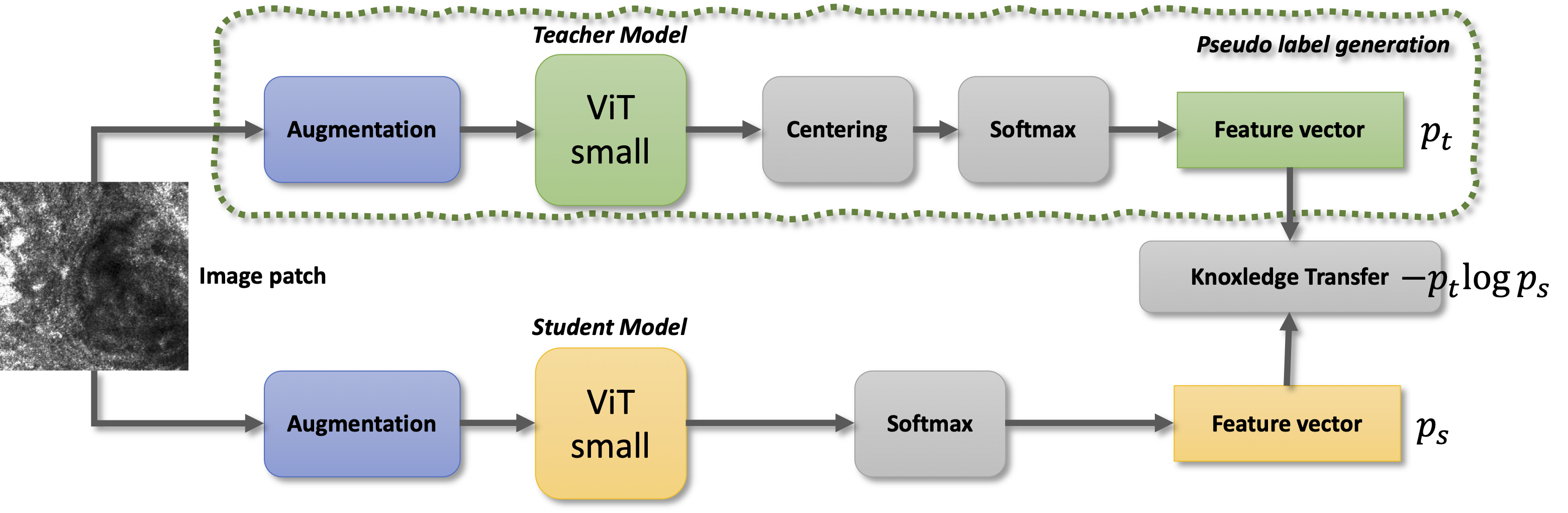}
    \caption{ViT small feature extraction model trained using the image patches from RCM images with the DINO algorithm.}
    \label{fig_dino}
\end{figure}

In our study, we employed a cutting-edge self-supervised learning algorithm known as DINO (self-distillation with no labels) \cite{caron2021emerging}, to effectively train a Vision Transformer (ViT) model \cite{dosovitskiy2020image} from scratch for the purpose of extracting features from image patches from RCM images within our dataset. The DINO algorithm is a self-supervised learning technique that uses knowledge distillation, where a student network learns to mimic the output of a teacher network, to train vision models without labeled data, enabling them to capture rich, high-level representations of images. Its unique approach leverages knowledge distillation techniques in an unsupervised setting, allowing the model to learn robust and meaningful representations of the data without the need for labeled training samples. In our study, we employed the ViT small model to extract a feature embedding consisting of 768 numbers. The DINO algorithm is illustrated in Figure \ref{fig_dino}.

\subsection{Clustering}


Following the feature extraction process, we employed the well-regarded $k$-means clustering algorithm \cite{hartigan1979algorithm} in statistical data analysis to categorize the image patches. The essence of $k$-means clustering is to partition the data into distinct groups (clusters) by assigning each data point to the cluster with the nearest mean, serving as a prototype. This method effectively grouped the image patches based on their extracted features, revealing patterns and similarities that may not be immediately apparent.

To determine the most effective number of clusters, we utilized the silhouette score \cite{gat2003scoring} as a guiding metric. The silhouette score measures the similarity of an object to its own cluster (cohesion) compared to other clusters (separation). By analyzing the silhouette scores for different values of $k$, we were able to identify the optimal number of clusters that best represented the inherent structure of our data, ensuring a robust and meaningful clustering outcome.

In order to add a layer of clinical relevance and validity to our computational findings, expert dermatologists (AW, JL) were actively involved in the analysis process. These experts meticulously examined the characteristics and clinical implications of each identified cluster. This comprehensive approach, integrating advanced AI algorithms with expert medical knowledge, underscores the multidisciplinary nature of our study.



\section{Results}

\begin{figure}
    \centering
    \includegraphics[width=0.6\linewidth]{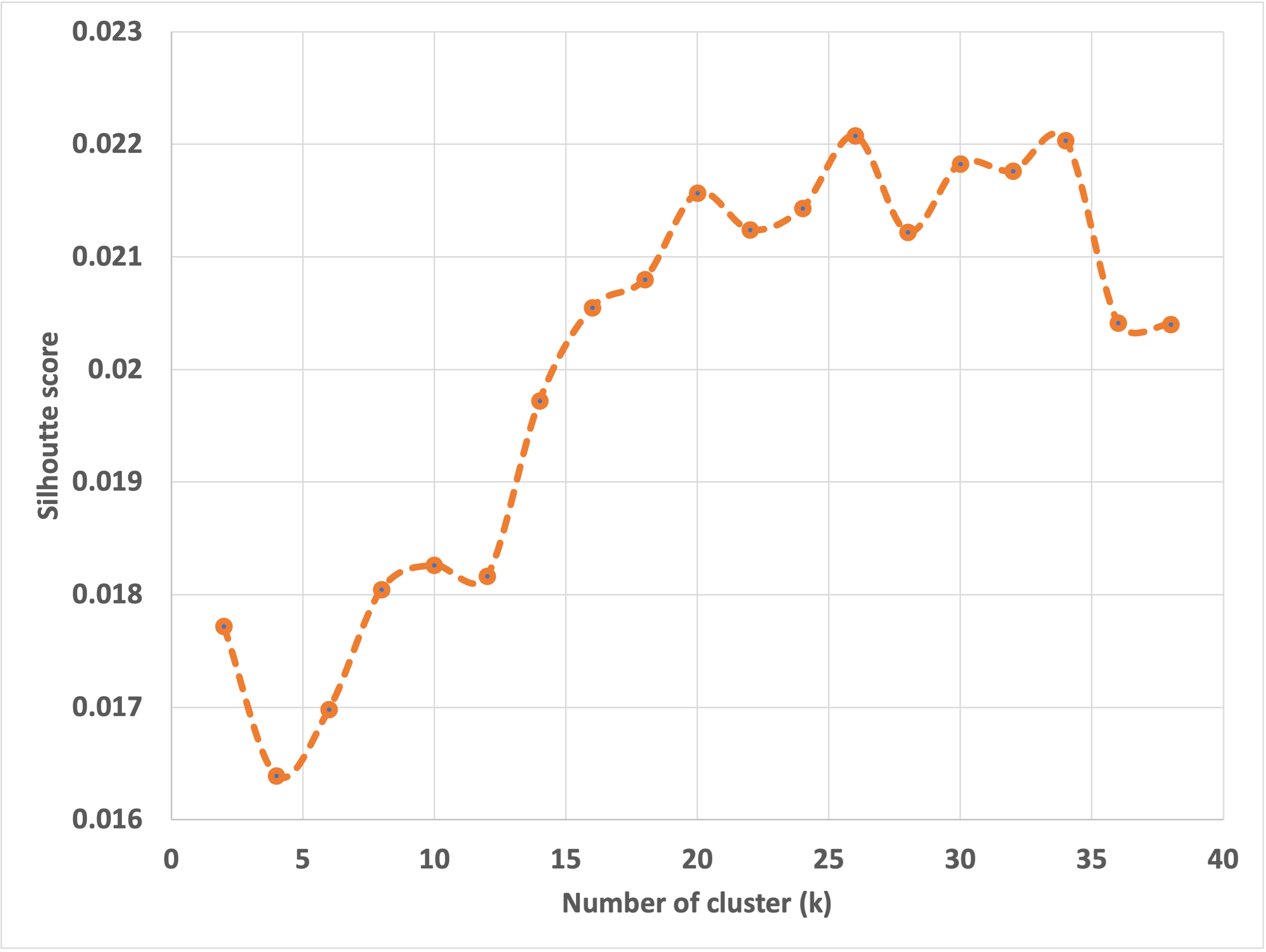}
    \caption{The silhouette scores of the clusters obtained from the image patch features using the $k$-means clustering algorithm are displayed. The x-axis represents the number of clusters $k$, while the y-axis denotes the silhouette score corresponding to each $k$.}
    \label{fig_silhoutte}
\end{figure}

The ViT small model was trained using the DINO algorithm in an unsupervised manner on the Summit supercomputer at the Oak Ridge Leadership Computing Facility (OLCF). The image patches were then used to extract features, which were clustered using the $k$-means clustering algorithm. Figure \ref{fig_silhoutte} illustrates the silhouette scores of the clusters determined by the $k$-means clustering algorithm for varying values of the number of clusters, $k$. Despite obtaining the highest silhouette score at $k=26$, we note a remarkable similarity in the silhouette scores from $k=18$ to $34$. Consequently, we determine the optimal number of clusters to be $k=18$, as it marks the inflection point where the trend shifts.

\begin{figure}
    \centering
    \begin{subfigure}{0.15\textwidth}
        \centering
        \includegraphics[width=0.95\linewidth]{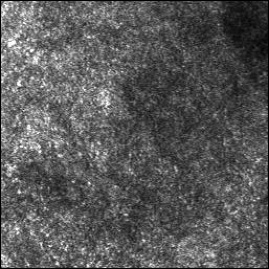}
        \caption{Cluster 0}
    \end{subfigure}
    \begin{subfigure}{0.15\textwidth}
        \centering
        \includegraphics[width=0.95\linewidth]{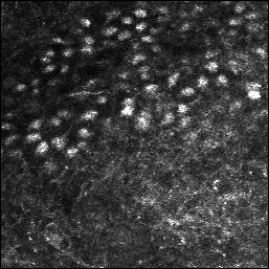}
        \caption{Cluster 1}
    \end{subfigure}
    \begin{subfigure}{0.15\textwidth}
        \centering
        \includegraphics[width=0.95\linewidth]{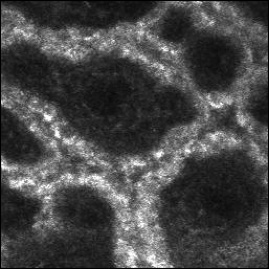}
        \caption{Cluster 2}
    \end{subfigure}
    \begin{subfigure}{0.15\textwidth}
        \centering
        \includegraphics[width=0.95\linewidth]{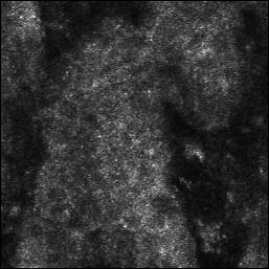}
        \caption{Cluster 3}
    \end{subfigure}
    \begin{subfigure}{0.15\textwidth}
        \centering
        \includegraphics[width=0.95\linewidth]{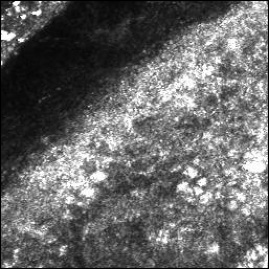}
        \caption{Cluster 4}
    \end{subfigure}
    \begin{subfigure}{0.15\textwidth}
        \centering
        \includegraphics[width=0.95\linewidth]{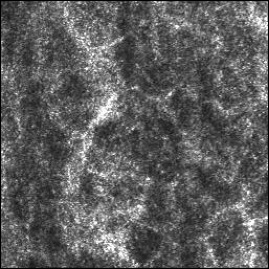}
        \caption{Cluster 5}
    \end{subfigure}

    \begin{subfigure}{0.15\textwidth}
        \centering
        \includegraphics[width=0.95\linewidth]{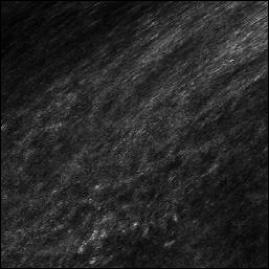}
        \caption{Cluster 6}
    \end{subfigure}
    \begin{subfigure}{0.15\textwidth}
        \centering
        \includegraphics[width=0.95\linewidth]{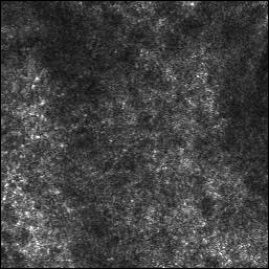}
        \caption{Cluster 7}
    \end{subfigure}
    \begin{subfigure}{0.15\textwidth}
        \centering
        \includegraphics[width=0.95\linewidth]{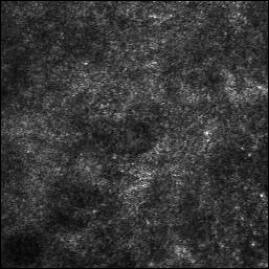}
        \caption{Cluster 8}
    \end{subfigure}
    \begin{subfigure}{0.15\textwidth}
        \centering
        \includegraphics[width=0.95\linewidth]{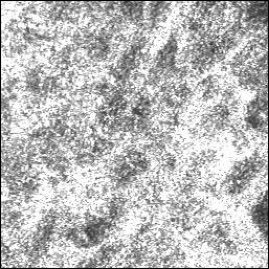}
        \caption{Cluster 9}
    \end{subfigure}
    \begin{subfigure}{0.15\textwidth}
        \centering
        \includegraphics[width=0.95\linewidth]{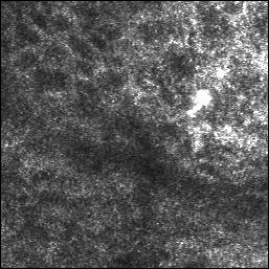}
        \caption{Cluster 10}
    \end{subfigure}
    \begin{subfigure}{0.15\textwidth}
        \centering
        \includegraphics[width=0.95\linewidth]{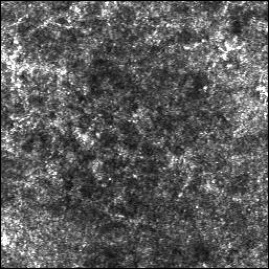}
        \caption{Cluster 11}
    \end{subfigure}

    \begin{subfigure}{0.15\textwidth}
        \centering
        \includegraphics[width=0.95\linewidth]{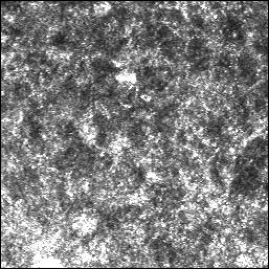}
        \caption{Cluster 12}
    \end{subfigure}
    \begin{subfigure}{0.15\textwidth}
        \centering
        \includegraphics[width=0.95\linewidth]{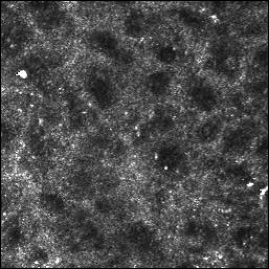}
        \caption{Cluster 13}
    \end{subfigure}
    \begin{subfigure}{0.15\textwidth}
        \centering
        \includegraphics[width=0.95\linewidth]{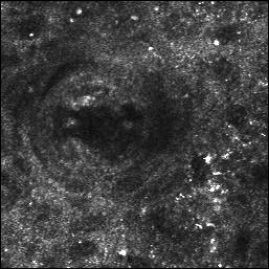}
        \caption{Cluster 14}
    \end{subfigure}
    \begin{subfigure}{0.15\textwidth}
        \centering
        \includegraphics[width=0.95\linewidth]{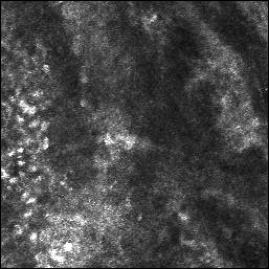}
        \caption{Cluster 15}
    \end{subfigure}
    \begin{subfigure}{0.15\textwidth}
        \centering
        \includegraphics[width=0.95\linewidth]{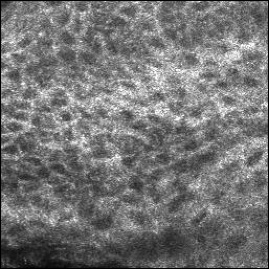}
        \caption{Cluster 16}
    \end{subfigure}
    \begin{subfigure}{0.15\textwidth}
        \centering
        \includegraphics[width=0.95\linewidth]{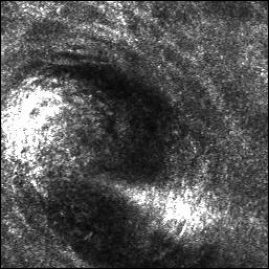}
        \caption{Cluster 17}
    \end{subfigure}

\caption{Example of patch images clustered by their image features determined by the ViT model trained by the DINO algorithm, associated with (a) cluster 0, (b) cluster 1, (c) cluster 2, and up to (r) cluster 17.}
\label{fig1}
\end{figure}

Two expert dermatologists (AW, JL), who have undergone extensive training, conducted a rigorous analysis of the patch images associated with each cluster. They meticulously identified the distinct characteristics and clinical implications of each cluster, as documented in Table \ref{table1clusterdesc}. Furthermore, the severity of each cluster is visually represented using color codes: green represents regular epidermis, yellow and orange represent irregular epidermis, red represents atypical cells, and blue represents artifacts. In order to provide a visual reference, Figure \ref{fig1} showcases sample images for each cluster.

\begin{table}[]
\centering
\caption{Clinical implications hold from the image patches in clusters determined by an expert dermatologist with the color codes: green (regular epidermis), yellow and orange (irregular epidermis), red (atypical cells), and blue (artifacts).}\label{table1clusterdesc}
\begin{tabular}{clcclc}
\hline
\multicolumn{1}{l}{Cluster} & Description & Color code & \multicolumn{1}{l}{Cluster} & Description & Color code \\ \hline
0 & Regular epidermis & Green & 9 & Regular epidermis & Green \\
1 & Inflammatory cells & Green & 10 & Epidermal invaginations & Green \\
2 & Bulbous projections & Green & 11 & \begin{tabular}{@{}c@{}}Epidermis streaming \\ with dentritic cells \end{tabular}& Orange \\
3 & \begin{tabular}{@{}c@{}}Bulbous projections, \\ epidermal invaginations\end{tabular} & Green & 12 & Pagetoid spread & Red \\
4 & Artifacts & Blue & 13 & Irregular epidermis type 4 & Yellow \\
5 & Irregular epidermis type 1 & Orange & 14 & \begin{tabular}{@{}c@{}}Meshwork pattern \\ (moderate atypia)\end{tabular} & Red \\
6 & Bundles of atypical cells & Red & 15 & \begin{tabular}{@{}c@{}}Meshwork pattern \\ (severe atypia)\end{tabular} & Red \\
7 & Irregular epidermis type 2 & Yellow & 16 & Irregular epidermis type 5 & Yellow \\
8 & Irregular epidermis type 3 & Yellow & 17 & Regular epidermis & Green \\ \hline
\end{tabular}
\end{table}


\begin{figure}
    \centering
    \begin{subfigure}{0.47\textwidth}
        \centering
        \includegraphics[width=0.97\linewidth]{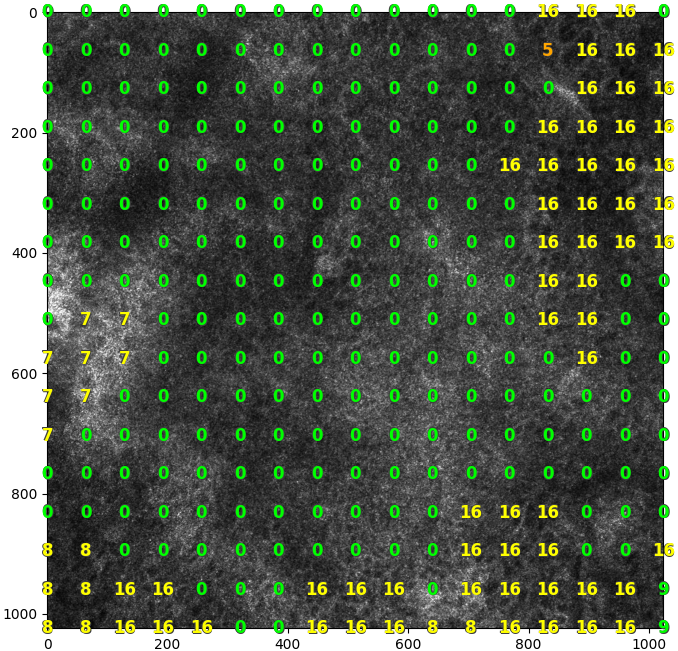}
        \caption{}
        \label{map_a}
    \end{subfigure}
    \begin{subfigure}{0.47\textwidth}
        \centering
        \includegraphics[width=0.95\linewidth]{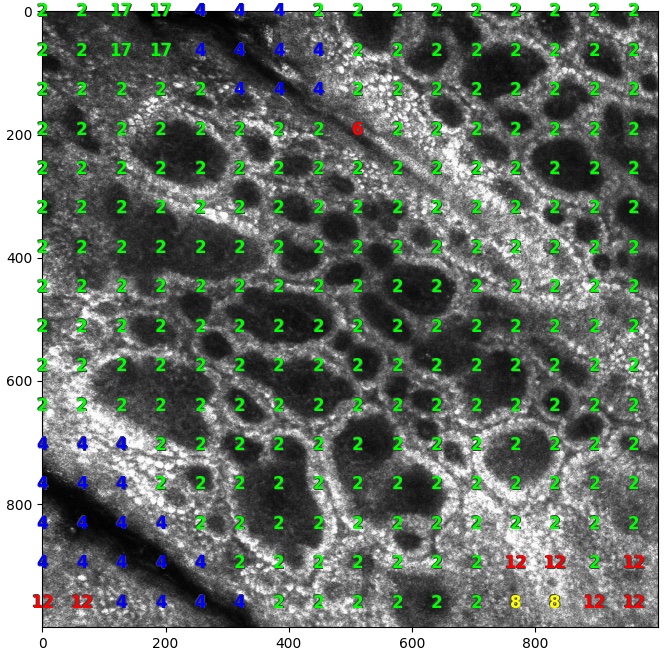}
        \caption{}
        \label{map_b}
    \end{subfigure}
    \begin{subfigure}{0.47\textwidth}
        \centering
        \includegraphics[width=0.95\linewidth]{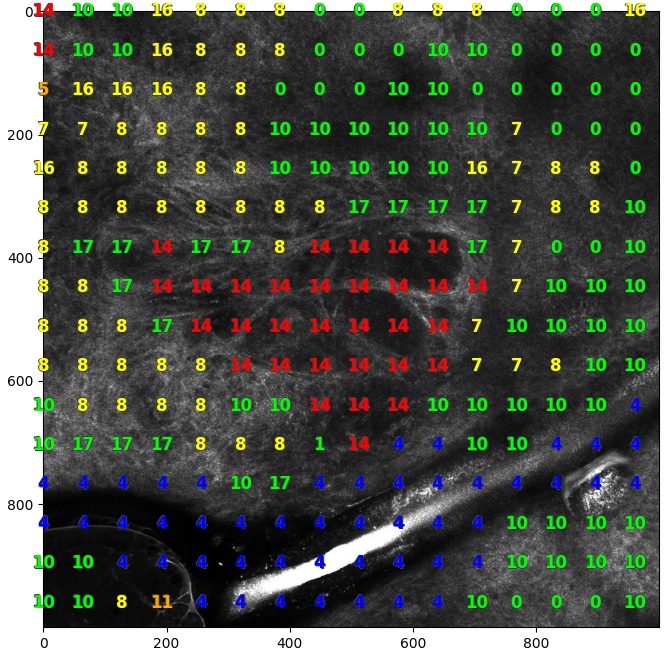}
        \caption{}
        \label{map_c}
    \end{subfigure}
    \begin{subfigure}{0.47\textwidth}
        \centering
        \includegraphics[width=0.97\linewidth]{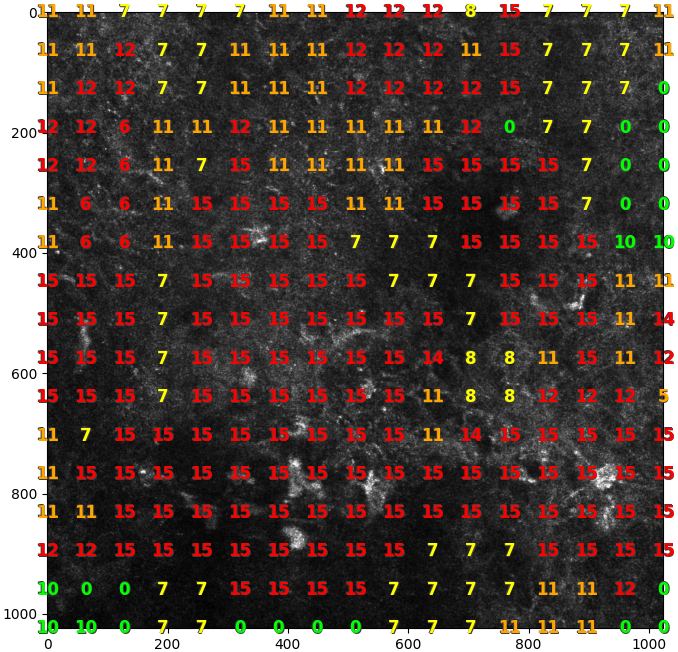}
        \caption{}
        \label{map_d}
    \end{subfigure}
    \caption{Sample RCM images are superimposed with the cluster map. (a) and (b) biopsy not recommended and (c) and (d) biopsy recommended cases.}
    \label{fig2}
\end{figure}

Figure \ref{fig2} displays a set of sample RCM images overlaid with cluster maps. The cluster maps indicate the class numbers and their associated risk levels using color codes. It is important to note that the images showcase the segmentation of cluster map regions based on their clinical characteristics, thereby aiding dermatologists in reading the cases more effectively and confidently diagnosing skin conditions.


\section{Discussion}
We conducted a proof-of-concept study that combined the latest advancements in AI and machine learning techniques with the clinical expertise of expert clinicians. This interdisciplinary collaboration yielded highly promising outcomes that can significantly advance the field of cancer diagnosis analytics and visualizations. Specifically, our study focused on automatically segmenting regions and visualizing severity in RCM images, and the results were remarkably effective.

For feature extraction, we chose to use the ViT small model, which proved to be a compelling approach. The combination of the ViT small model and the DINO algorithm for training demonstrated its effectiveness in our study. However, further comparative studies are needed to achieve a more comprehensive representation of image features. For example, the ViT models offer three variations (small, base, and large) based on the size of their trainable parameters. Additionally, exploring popular convolutional neural network (CNN) models \cite{krizhevsky2012imagenet,he2016deep,simonyan2014very,huang2017densely}, known for their ability to infer image features rapidly and with lower computational power, could be a valuable avenue for future research. Furthermore, considering other self-supervised learning and unsupervised learning techniques, such as bootstrap your own latent (BYOL)\cite{grill2020bootstrap}, presents intriguing possibilities for follow-up investigations.

While the choice of using $k$-means clustering was commendable, further studies comparing different clustering algorithms\cite{ester1996density,reynolds2009gaussian,zhang1996birch,mullner2011modern} are necessary to gain deeper insights into their performance. Evaluating these algorithms, along with selecting optimal feature extraction models, can be done using the silhouette score as a performance metric. A higher score indicates superior performance. It is important to note that the order of clusters and the association of samples may vary during the clustering algorithm, requiring repetitive evaluation and interpretation of the clinical significance associated with each cluster by expert clinicians if we repeat the clustering process. Since this step is time-consuming and demands significant human effort, minimizing this effort is crucial for achieving study efficiency.

The findings depicted in Figure \ref{fig2} underscore the efficacy of our algorithm and system in automatically segmenting and determining the severity of abnormalities in RCM images. It is important to highlight that the presence of atypical regions identified by the algorithm does not automatically imply the necessity of a biopsy, as exemplified in Figure \ref{map_b}. This could be attributed to either misidentification of certain areas or erroneous readings. Furthermore, it is noteworthy that these abnormal regions may not encompass the entirety of the image but can still occupy a considerable portion, as demonstrated in Figure \ref{map_c}.

Looking ahead, a promising avenue for further scientific investigation and development lies in the utilization of cluster maps in clinical decision-making. For instance, quantifying the extent of connected atypical regions and establishing a threshold could potentially serve as a criterion for clinical judgments. Nonetheless, it is crucial to acknowledge that such research would necessitate a more extensive dataset that accurately represents the entire disease spectrum in order to draw robust and dependable conclusions.

\section{Conclusion}

In this study, we utilized a self-supervised training algorithm, DINO, to develop an ML model that can extract local textual features from image patches of RCM images. We then grouped the images based on their features using the k-means clustering algorithm. 
Our results suggest that unsupervised clustering of the features extracted from the images can be used for downstream clinical classification tasks.

Our research indicates that the feature map has clinical relevance and that the cluster map can be utilized as a segmentation tool for the RCM images. This can assist dermatologists in reading cases more effectively and confidently diagnosing skin conditions. Furthermore, our approach could be utilized as an educational tool for dermatologic trainees.

\section*{Acknowledgments}
This research used resources of the Oak Ridge Leadership Computing Facility at the Oak Ridge National Laboratory, which is supported by the DOE Office of Science under Contract No. DE-AC05-00OR22725.

The study was supported by the Seed Money Fund program of Oak Ridge National Laboratory, under the Laboratory Directed Research and Development (LDRD) No. 11013.

\bibliographystyle{unsrt}  
\bibliography{references}

\end{document}